\documentclass{article}

\usepackage{arxiv}

\usepackage[utf8]{inputenc} 
\usepackage[T1]{fontenc}    
\usepackage{hyperref}       
\usepackage{url}            
\usepackage{booktabs}       
\usepackage{amsfonts}       
\usepackage{nicefrac}       
\usepackage{microtype}      
\usepackage{lipsum}		
\usepackage{graphicx}
\usepackage{natbib}
\usepackage{doi}
\setcitestyle{numbers}

\usepackage{booktabs} 
\usepackage{algorithmic}
\usepackage{amsmath,amssymb,amsfonts}
\usepackage[table,xcdraw]{xcolor}
\usepackage{epsfig}
\usepackage{multirow}
\usepackage{makecell}
\usepackage[caption=false]{subfig}
\usepackage{diagbox}
\usepackage{adjustbox}
\usepackage{balance}
\usepackage{tcolorbox}
\usepackage{hyperref}
\usepackage{paralist}
\usepackage{tikz-cd}
\usepackage{soul}
\usepackage{threeparttable}

\usepackage{pgfplots}
\usepackage{ifthen}             
\usepackage{tikz}
\usetikzlibrary{topaths,calc}

\def\rotateclockwise#1{
  \newdimen\xrw
  \pgfextractx{\xrw}{#1}
  \newdimen\yrw
  \pgfextracty{\yrw}{#1}
  \pgfpoint{\yrw}{-\xrw}
}

\def\rotatecounterclockwise#1{
  \newdimen\xrcw
  \pgfextractx{\xrcw}{#1}
  \newdimen\yrcw
  \pgfextracty{\yrcw}{#1}
  \pgfpoint{-\yrcw}{\xrcw}
}

\def\outsidespacerpgfclockwise#1#2#3{
  \pgfpointscale{#3}{
    \rotateclockwise{
      \pgfpointnormalised{
        \pgfpointdiff{#1}{#2}}}}
}

\def\outsidespacerpgfcounterclockwise#1#2#3{
  \pgfpointscale{#3}{
    \rotatecounterclockwise{
      \pgfpointnormalised{
        \pgfpointdiff{#1}{#2}}}}
}

\def\outsidepgfclockwise#1#2#3{
  \pgfpointadd{#2}{\outsidespacerpgfclockwise{#1}{#2}{#3}}
}

\def\outsidepgfcounterclockwise#1#2#3{
  \pgfpointadd{#2}{\outsidespacerpgfcounterclockwise{#1}{#2}{#3}}
}

\def\outside#1#2#3{
  ($ (#2) ! #3 ! -90 : (#1) $)
}

\def\cornerpgf#1#2#3#4{
  \pgfextra{
    \pgfmathanglebetweenpoints{#2}{\outsidepgfcounterclockwise{#1}{#2}{#4}}
    \let\anglea\pgfmathresult
    \let\startangle\pgfmathresult

    \pgfmathanglebetweenpoints{#2}{\outsidepgfclockwise{#3}{#2}{#4}}
    \pgfmathparse{\pgfmathresult - \anglea}
    \pgfmathroundto{\pgfmathresult}
    \let\arcangle\pgfmathresult
    \ifthenelse{180=\arcangle \or 180<\arcangle}{
      \pgfmathparse{-360 + \arcangle}}{
      \pgfmathparse{\arcangle}}
    \let\deltaangle\pgfmathresult

    \newdimen\x
    \pgfextractx{\x}{\outsidepgfcounterclockwise{#1}{#2}{#4}}
    \newdimen\y
    \pgfextracty{\y}{\outsidepgfcounterclockwise{#1}{#2}{#4}}
  }
  -- (\x,\y) arc [start angle=\startangle, delta angle=\deltaangle, radius=#4]
}

\def\corner#1#2#3#4{
  \cornerpgf{\pgfpointanchor{#1}{center}}{\pgfpointanchor{#2}{center}}{\pgfpointanchor{#3}{center}}{#4}
}

\def\hedgem#1#2#3#4{
  
  \outside{#1}{#2}{#4}
  \pgfextra{
    \def\hgnodea{#1}
    \def\hgnodeb{#2}
  }
  foreach \c in {#3} {
    \corner{\hgnodea}{\hgnodeb}{\c}{#4}
    \pgfextra{
      \global\let\hgnodea\hgnodeb
      \global\let\hgnodeb\c
    }
  }
  \corner{\hgnodea}{\hgnodeb}{#1}{#4}
  \corner{\hgnodeb}{#1}{#2}{#4}
  -- cycle
}

\newtheorem{definition}{Definition}[section]
\newcommand{\curlyL}{\mathcal{L}}
\DeclareMathOperator{\Ima}{Im}

\title{Higher-Order Networks\\ Representation and Learning: A Survey}

\date{} 	

\author{ 
    Hao Tian and Reza Zafarani\\
	Data Lab, EECS Department\\
	Syracuse University\\
	\texttt{\{haotian,reza\}@data.syr.edu} \\
}

\begin{document}
\maketitle

\begin{abstract}
	Network data has become widespread, larger, and more complex over the years. Traditional network data is \textit{dyadic}, capturing the relations among \underline{pairs} of entities. With the need to model interactions among more than two entities, significant research has focused on \textit{higher-order networks} and ways to represent, analyze, and learn from them. There are two main directions to studying higher-order networks. One direction has focused on capturing \textit{higher-order patterns} in traditional (dyadic) graphs by changing the basic unit of study from nodes to small frequently observed subgraphs, called \textit{motifs}. As most existing network data comes in the form of pairwise dyadic relationships, studying higher-order structures within such graphs may uncover new insights. The second direction aims to directly model higher-order interactions using new and more complex representations such as \textit{simplicial complexes} or \textit{hypergraphs}. Some of these models have long been proposed, but improvements in computational power and the advent of new computational techniques have increased their popularity. Our goal in this paper is to provide a succinct yet comprehensive summary of the advanced higher-order network analysis techniques. We provide a systematic review of its foundations and algorithms, along with use cases and applications of higher-order networks in various scientific domains. 
\end{abstract}


\section{Introduction}
Networks are natural representations of relationships between entities using nodes and edges~\cite{barabasi2016network}. Real-world networks are observed everywhere: the structure of chemical substances, ecological systems, communication networks, air and land transportation networks, power grids, among many other examples. Many problems can be naturally described and solved using networks. For instance, epidemic models on networks~\cite{HARKO2014184} can help predict the spread of pandemics by analyzing the interaction network among infected individuals; link-analysis methods such as PageRank~\cite{ilprints422} can help assess the importance of Websites, which in turn can be used to fine-tune searching engine results; Shortest paths algorithms~\cite{DBLP:journals/corr/MadkourARRB17} calculate the most efficient driving route between two locations on the transportation networks. With the rise in demand to model systems with more complex information, researchers have enriched network models by adding additional attributes to nodes and edges: weights, signs, labels, timestamps, and even metadata. Some models have even changed or extended the network basics. An example is a \textit{heterogeneous network}~\cite{DBLP:journals/corr/abs-2004-00216}, where nodes and edges can be of different types. Another example is a \textit{dynamic network}~\cite{10.1145/3172867}, where each node or edge can exist only for a specific period of time. While models for networks have been enriched from various aspects, in most network models, edges still represent \textit{dyadic} relationships, that is, relationships among two entities.\vspace{1mm} 

Dyadic relationships are insufficient in many real-world scenarios, specifically when there is an interaction involving more than two entities. For example, a social event may include more than two people. This is not equivalent to social interactions among all pairs of people in the event. However, such a \textit{higher-order} pattern frequently occurs in social networks due to \textit{triadic closure}~\cite{Granovetter:1973p66}, where a triangle's formation is often dependent on three edges. Another example is the frequent appearance of some specific small subgraphs (with more than two nodes) in real-world networks~\cite{alex:hal-03266728, leinhardt1971structure, DBLP:journals/corr/abs-1304-1548}. In 2002, Shen-Orr et al.~\cite{ShenOrr2002NetworkMI} introduced the term \textit{network motifs} to represent such frequent subgraphs as the building blocks of transcriptional regulation
networks. The idea was further explored for various types of graphs by Milo et al.~\cite{alon-network-motifs}, where they showed that different types of networks can be distinguished using motif counts as features~\cite{doi:10.1126/science.1089167}. Such discoveries clearly indicate that \textit{it is insufficient to only model dyadic networks}. 

As a result, modeling higher-order networks has a long history. Some higher-order representations are proposed as early as the 1960s-1970s, e.g., \textit{simplicial complexes}~\cite{115026}. However, there were many obstacles to utilizing such higher-order representations at that time. First, the computational power was insufficient to compute using such representations, even for counting simple motifs. Second, without the development of the internet, data collection was inefficient and small-scale. Hence, there was an earlier decrease in the demand to model and represent higher-order networks.

\begin{figure*}[ht]
    \centering
	\subfloat[\small{Simple graph and its corresponding motif graph}]{%
		\resizebox{.65\linewidth}{!}{\begin{tikzpicture}
\draw[ultra thin, white] (0,0) -- (4.5,0) -- (4.5,3.5) -- (0,3.5) -- (0,0);

\coordinate (a) at (1.5, 2.5);
\coordinate (b) at (0.5, 1.5);
\coordinate (c) at (1.5, 0.5);
\coordinate (d) at (2.5, 1.5);
\coordinate (e) at (3.5, 1.5);

\draw[very thin] (a) -- (c) -- (b) -- (a) -- (d) -- (c);
\draw[very thin] (d) -- (e);

\draw[fill=black] (a) circle (3pt);
\draw[fill=black] (b) circle (3pt);
\draw[fill=black] (c) circle (3pt);
\draw[fill=black] (d) circle (3pt);
\draw[fill=black] (e) circle (3pt);

\node[font=\large] at (1.5, 3) {$v_1$};
\node[font=\large] at (0.5, 2) {$v_2$};
\node[font=\large] at (1.5, 0) {$v_3$};
\node[font=\large] at (2.5, 2) {$v_4$};
\node[font=\large] at (3.5, 2) {$v_5$};

\draw[ultra thin, white] (6+0,0) -- (6+4.5,0) -- (6+4.5,3.5) -- (6+0,3.5) -- (6+0,0);
\coordinate (a) at (6+1.5, 2.5);
\coordinate (b) at (6+0.5, 1.5);
\coordinate (c) at (6+1.5, 0.5);
\coordinate (d) at (6+2.5, 1.5);
\coordinate (e) at (6+3.5, 1.5);

\filldraw[very thin, draw=black, fill=blue!20] (a) -- (b) -- (c) -- cycle;
\filldraw[very thin, draw=black, fill=blue!20] (a) -- (d) -- (c) -- cycle;

\draw[fill=black] (a) circle (3pt);
\draw[fill=black] (b) circle (3pt);
\draw[fill=black] (c) circle (3pt);
\draw[fill=black] (d) circle (3pt);
\draw[fill=black] (e) circle (3pt);

\node[font=\large] at (6+1.5, 3) {$v_1$};
\node[font=\large] at (6+0.5, 2) {$v_2$};
\node[font=\large] at (6+1.5, 0) {$v_3$};
\node[font=\large] at (6+2.5, 2) {$v_4$};
\node[font=\large] at (6+3.5, 2) {$v_5$};

\node[font=\large] at (6+1, 1.5) {$M_1$};
\node[font=\large] at (6+2, 1.5) {$M_2$};

\node[font=\large] at (6+1, 2.5) {1};
\node[font=\large] at (6+1, 0.5) {1};
\node[font=\large] at (6+2, 2.5) {1};
\node[font=\large] at (6+2, 0.5) {1};
\node[font=\large] at (6+1.3, 1.9) {2};

\end{tikzpicture}}%
	}\hfill
	\subfloat[\small{Simplicial complex}]{%
		\resizebox{.36\linewidth}{!}{\begin{tikzpicture}

\draw[ultra thin, white] (0,0) -- (4,0) -- (4,4) -- (0,4) -- (0,0);
\coordinate (a) at (0.5+0.5, 0.5+1.5);
\coordinate (b) at (0.5+0.5, 0.5+2.5);
\coordinate (c) at (0.5+1.5, 0.5+2);
\coordinate (d) at (0.5+2.4, 0.5+2.8);
\coordinate (e) at (0.5+2.9, 0.5+2);
\coordinate (f) at (0.5+2.4, 0.5+1.2);
\coordinate (g) at (0.5+1, 0.5+0.5);

\filldraw[very thin, draw=black, fill=blue!20] (a) -- (b) -- (c) -- cycle;
\filldraw[very thin, draw=black, fill=blue!20] (c) -- (d) -- (e) -- (f) -- cycle;
\draw[very thin, dashed] (c) -- (e);
\draw[very thin] (d) -- (f);

\fill (a) circle (0.08) node [below left] {$v_1$};
\fill (b) circle (0.08) node [above left] {$v_2$};
\fill (c) circle (0.08) node [below] {$v_3$};
\fill (d) circle (0.08) node [above] {$v_5$};
\fill (e) circle (0.08) node [right] {$v_6$};
\fill (f) circle (0.08) node [below] {$v_7$};
\fill (g) circle (0.08) node [above] {$v_4$};

\end{tikzpicture}}%
	}\hfill
	\subfloat[\small{Hypergraph}]{%
		\resizebox{.36\linewidth}{!}{\begin{tikzpicture}
    \node (v1) at (0,2) {};
    \node (v2) at (1.5,3) {};
    \node (v3) at (4,2.5) {};
    \node (v4) at (0,0) {};
    \node (v5) at (2,0.5) {};
    \node (v6) at (2.8,0.8) {};
    \node (v7) at (3.5,0) {};

    \begin{scope}[fill opacity=0.8]
    \filldraw[fill=yellow!70] ($(v1)+(-0.5,0)$) 
        to[out=90,in=180] ($(v2) + (0,0.5)$) 
        to[out=0,in=90] ($(v3) + (1,0)$)
        to[out=270,in=0] ($(v2) + (1,-0.8)$)
        to[out=180,in=270] ($(v1)+(-0.5,0)$);
    \filldraw[fill=blue!70] ($(v4)+(-0.5,0.2)$)
        to[out=90,in=180] ($(v4)+(0,1)$)
        to[out=0,in=90] ($(v4)+(0.6,0.3)$)
        to[out=270,in=0] ($(v4)+(0,-0.6)$)
        to[out=180,in=270] ($(v4)+(-0.5,0.2)$);
    \filldraw[fill=green!80] ($(v5)+(-0.5,0)$)
        to[out=90,in=225] ($(v3)+(-0.5,-1)$)
        to[out=45,in=270] ($(v3)+(-0.7,0)$)
        to[out=90,in=180] ($(v3)+(0,0.5)$)
        to[out=0,in=90] ($(v3)+(0.7,0)$)
        to[out=270,in=90] ($(v3)+(-0.3,-1.8)$)
        to[out=270,in=90] ($(v7)+(0.5,-0.3)$)
        to[out=270,in=270] ($(v5)+(-0.5,0)$);
    \filldraw[fill=red!70] ($(v2)+(-0.5,-0.2)$) 
        to[out=90,in=180] ($(v2) + (0.2,0.4)$) 
        to[out=0,in=180] ($(v3) + (0,0.3)$)
        to[out=0,in=90] ($(v3) + (0.3,-0.1)$)
        to[out=270,in=0] ($(v3) + (0,-0.3)$)
        to[out=180,in=0] ($(v3) + (-1.3,0)$)
        to[out=180,in=270] ($(v2)+(-0.5,-0.2)$);
    \end{scope}

    \foreach \v in {1,2,...,7} {
        \fill (v\v) circle (0.1);
    }

    \fill (v1) circle (0.1) node [right] {$v_1$};
    \fill (v2) circle (0.1) node [below left] {$v_2$};
    \fill (v3) circle (0.1) node [left] {$v_3$};
    \fill (v4) circle (0.1) node [below] {$v_4$};
    \fill (v5) circle (0.1) node [below right] {$v_5$};
    \fill (v6) circle (0.1) node [below right] {$v_6$};
    \fill (v7) circle (0.1) node [below left] {$v_7$};

\end{tikzpicture}}%
	}\vspace{-1mm}
	\caption{Comparison among Higher-Order Network Representations. (a). A simple graph (left) along with its \textit{motif graph} (right) of \textit{3-Cliques} (triangles). With the same set of nodes, motif graphs transform edges into ``membership in given motifs." From the example, the given motif is a triangle, so there are two triangles detected and only one edge $(v_1,v_3)$ shared by both triangles. The edge $(v_4,v_4)$ that does not belong to any motif will be ignored. (b). A Simplicial Complex, including a 0-simplex (single node), a 2-simplex (triangle), and a 3-simplex (tetrahedron). Note that any \textit{face} (sub-simplex) of an existing simplex is also included in the simplicial complex, for example, $(v_5,v_6,v_7)$.  (c). A Hypergraph. Unlike simplicial complexes, any subedge of hyperedges does not have to appear in the edge set, i.e., $(v_1,v_2,v_3)$ and $(v_2,v_3)$ are two different hyperedges. }
	\label{fig1:compare_hons}
\end{figure*}
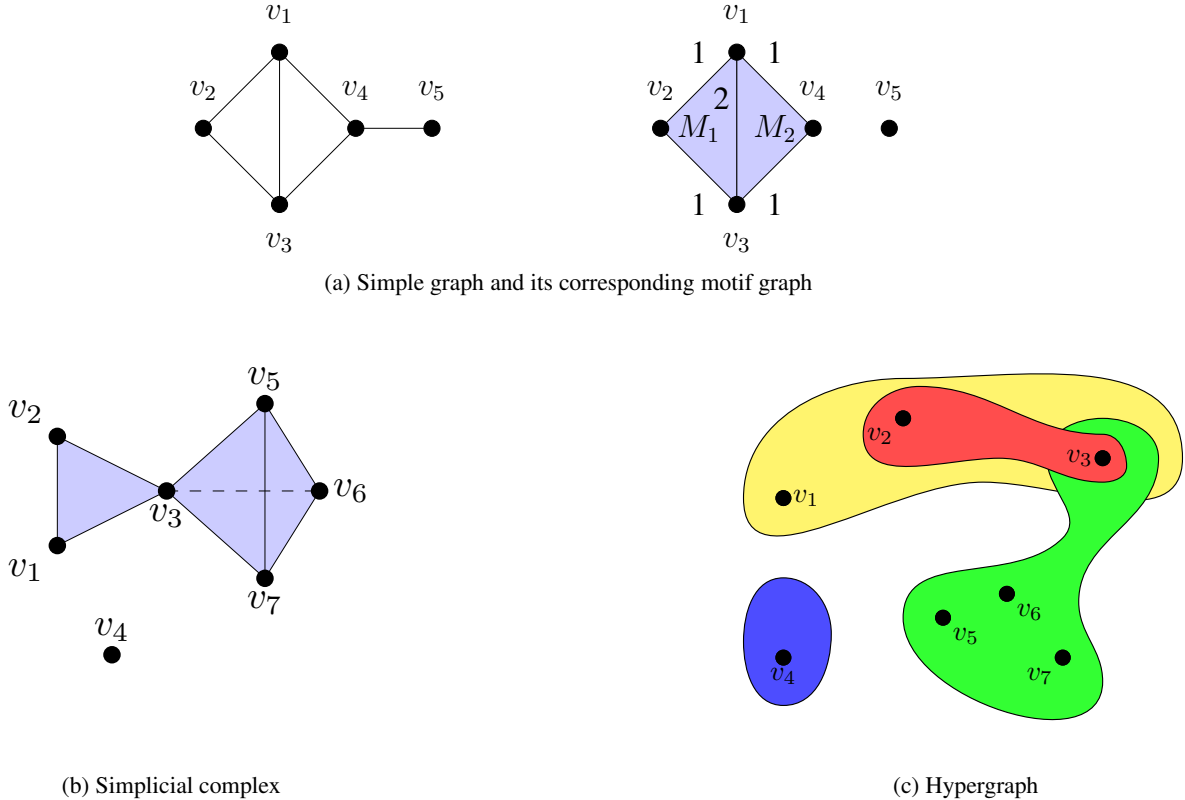

With the fast development of computational resources and algorithmic tools, higher-order network analysis is now widely used across various fields leading to various discoveries. For instance, in brain networks, some motifs with high functionality are generated more than other motifs to increase neural efficiency~\cite{brain_article,Duclos2020BrainNM}; In biological networks, motifs are commonly found and capture evolution~\cite{Kashtan13773}; In social networks, motifs help understand group interactions~\cite{leskovec2010signed,articlehong}. As most existing network data is already collected as dyadic graphs, it is often impossible to recover the original higher-order interactions (if they exist). In recent years, more higher-order network data is collected, e.g., in collaboration networks or on hashtags~\cite{DBLP:journals/corr/abs-1802-06916}. As a result, more complex higher-order algorithms and representations can now be directly applied. Our goal is to broadly survey such ways to represent, analyze, and learn from higher-order networks.\vspace{2mm}

\textbf{Higher-Order Network Representations:} there are three main branches of network abstractions that are widely studied or utilized for higher-order networks --- \textit{motifs}, \textit{simplicial complexes}, and \textit{hypergraphs}. Before diving into detailed mathematical representations of them, we briefly describe their differences using concrete examples involving three individuals \texttt{A}, \texttt{B}, and \texttt{C}~\cite{DBLP:journals/corr/abs-2103-05031}:
\begin{enumerate}
    \item \textbf{Network Motifs}: \texttt{A}, \texttt{B}, and \texttt{C} have contact information of \textbf{each other} in their contact list. They form a triangle (an example of a \textit{motif}). 
    \item \textbf{Simplicial Complexes}: \texttt{A}, \texttt{B}, and \texttt{C} are in the same class in high school. \textbf{Any subset of \{\texttt{A}, \texttt{B}, \texttt{C}\}} indicates a classmate relationship. \texttt{A}, \texttt{B}, and \texttt{C} form a \textit{simplex}, a basic unit of a \textit{simplicial complex}.
    \item \textbf{Hypergraphs}: \texttt{A}, \texttt{B}, and \texttt{C} publish a paper together. We create a new \textit{hyperedge} representing the collection of \textbf{all authors} of this paper, i.e., hyperedge \{\texttt{A}, \texttt{B}, \texttt{C}\}. \texttt{A}, \texttt{B}, and \texttt{C} form a \textit{hypergraph} with a single hyperedge. Note that \{\texttt{A}, \texttt{B}\},\{\texttt{A}, \texttt{C}\} and \{\texttt{B}, \texttt{C}\} are not included as hyperedges in the hypergraph. 
\end{enumerate}

Figure~\ref{fig1:compare_hons} shows a comparison of these network representations. We still often study motifs in dyadic graphs, but instead of looking at pairwise relationships, we extract higher-order relationships. For example, we can identify whether an edge is part of some prespecified motif and set edge weights to the number of times the edge belongs to such motif. As shown in Figure~\ref{fig1:compare_hons} (a), one has to specify a motif to study, for example, a triangle. Then the original graph can be transformed by only keeping its edges that belong to such a given motif. Figure~\ref{fig1:compare_hons} (b) and (c) show a simplicial complex and a hypergraph with similar structures. Both edges (simplexes) can be represented by sets. However, in simplicial complexes, any subset simplex, by definition, also exists, while in hypergraphs, a hyperedge only acknowledges the existence of the exact set. For example ($v_1$,$v_2$,$v_3$) and ($v_2$,$v_3$) exist in Figure~\ref{fig1:compare_hons} (c), while their subedges ($v_1$,$v_2$) and ($v_1$,$v_3$) do not exist.\vspace{1mm} 

\noindent \textbf{Topics not covered in this survey.} Some studies are outside of the scope of this paper and presenting them might obfuscate some of the formalism used in this survey. These studies either (1) apply a higher-order abstraction but outside the field of networks; or (2) use similar terminologies as those of higher-order networks but with a different meaning across various domains. Here, we list some such areas:
\begin{itemize}
    \item \textit{Network of Networks (NoN)}, or \textit{multilayer networks}~\cite{doi:10.1137/1.9781611975321.77,10.1093/comnet/cnu016, 7956273}, are basically heterogeneous networks. Such models combine networks from different sources and merge into a larger, more complex network. Here, nodes and edges can be different kinds of entities. The basic unit of such networks is one type of network, but not subgraphs. 
    \item \textit{Higher-Order Markov Models}: These studies investigate higher-order dependencies in random walks on networks~\cite{lambiotte2018understanding, MASUDA20171,doi:10.1126/sciadv.1600028}. In some cases, first-order random walks are not able to describe network flows, so it becomes necessary to utilize higher-order Markov chains to fit the real-world observations in networks. 
    \item \textit{Higher-order Graph Signal Processing}: Graph signal processing~\cite{8347162} has also been extended from dyadic networks to simplicial complexes~\cite{DBLP:journals/corr/abs-1210-4752, DBLP:journals/corr/abs-1211-0053,articlegsp}. In graph signal processing, vertices carry samples of signals, and edges capture linear transformations on such signals. Such a system is denoted by a \textit{graph filter}, which aims to model complex signal transformations based on a graph structure. For a comprehensive review of higher-order graph signal processing, refer to~\cite{SCHAUB2021108149}.
    \item \textit{Higher-order Dynamical Systems on Networks}: Network dynamical system models pairwise node interactions based on dynamic network structure~\cite{BOCCALETTI2006175}. For example, robots can form real-time shapes together by looking at their neighbors. Such pairwise interactions are extended to higher-order interactions~\cite{DBLP:journals/corr/abs-2104-11329}, using either simplicial complexes~\cite{Mill_n_2020} or hypergraphs~\cite{PhysRevE.101.062313}. 
\end{itemize}

\noindent \textbf{Scope and Organization.} Compared to other surveys, our goal is to provide a succinct yet broad and comprehensive survey that focuses on higher-order networks. Abstract network topologies are categorized on the basis of the data type and applications. The surveys aims to assist researchers in identifying the appropriate methods, resources, and higher-order tools for their research tasks.

The rest of this paper is structured as follows. Section~\ref{sec:preliminary} introduces necessary dyadic graph basics. Sections~\ref{sec:motif}, \ref{sec:simplicial_complexes}, and \ref{sec:hypergraph} introduce foundations, algorithms, and applications of network motifs, simplicial complexes, and hypergraphs, respectively. Section~\ref{sec:dataset_tool} summarizes existing datasets and tools that have been used in higher-order network studies. We present challenges and future direction and conclude in Section~\ref{sec:conclusion}.

\section{Preliminary of Graphs}
\label{sec:preliminary}

We briefly introduce some basic dyadic graph concepts, which are being applied and generalized in various higher-order representations. 

\begin{itemize}
    \item Undirected/Directed Graph: An \textit{undirected graph} is an ordered pair $G = (V,E)$. Set $V = \{v_1, v_2, ...,v_n\}$ is the set of vertices and set $E \subseteq \{\{v_i,v_j\}|v_i,v_j \in V \}$ is the set of edges, where $\{v_i,v_j\}$ is an unordered pair of nodes. In contrast, a \textit{directed graph} is an ordered pair $G = (V,E)$, where $V$ is the set of vertices and $E \subseteq \{(v_i,v_j)|(v_i,v_j) \in V^2 \}$ are ordered pairs of nodes. 
    
    \item Simple Graph: A \textit{self-loop} is an edge that starts and ends at the same vertex, for example, $(v_i,v_i)$. Duplicate edges in an edge set are called \textit{multiple edges}, \textit{multi-edge}, or \textit{parallel edges}. A graph without any self-loop or multiple edges is a \textit{simple graph}. Most graph-based studies focus on simple graphs.

    \item Weighted Graph: A \textit{weighted graph} $G=(V,E,w)$ is a graph with assigned weights $w: E\rightarrow \mathbb{R} $ to its edges. 
    
    \item Graph Isomorphism: Graphs $G$ and $H$ are \textit{isomorphic}, denoted as $G\simeq H$, if there is a bijection between the vertex sets of $G$ and $H$ $$f:V(G)\rightarrow V(H),$$
    such that any two vertices $u$ and $v$ of $G$ are adjacent in $G$ if and only if $f(u)$ and $f(v)$ are adjacent in $H$.

    \item Walk, Trail, Path and Cycle: A \textit{walk} is a sequence of vertices and edges of a graph. For example, one can traverse from one vertex to another once there is an edge between them. A walk is said to be \textit{open} when the starting and ending nodes are different, and \textit{closed}, otherwise. A \textit{trail} is an open walk where one edge is repeated. A \textit{path} is a trail in which neither a vertex nor an edge is repeated. A \textit{cycle} is a closed walk that neither a vertex nor an edge is repeated.
        
    \item Cut, Volume, and Conductance: A \textit{cut} is a partition of the vertices of a graph into two disjoint subsets, marked as $(S,\Bar{S})$. The \textit{volume} of a node set $S\subseteq V$ is defined as the total number (or weight) of the edges incident with $S$, denoted as $\text{vol}(S)$. The \textit{conductance} measures the `goodness' of a cut separating a graph, $$\varphi(S) = \frac{\sum_{i\in S,j\in \Bar{S}}A_{ij}}{\min(\text{vol}(S),\text{vol}(\Bar{S}))},$$ where $A_{ij}$ are the entries of the adjacency matrix for $G$. Lower conductance ensures a balanced cut with fewer cross-edges (between $S$ and $\Bar{S}$). 

    \item Adjacency Matrix: a square matrix used to represent a finite graph. Assume a graph has $n$ nodes. Its corresponding adjacency matrix $A$ is a matrix of size $n\times n$, where $A_{i,j}=1$ when nodes $i$ and $j$ are connected and $A_{i,j}=0$, otherwise. Adjacency matrices of undirected graphs are symmetric. Adjacency matrices of simple graphs are binary, with all zeros on the main diagonal. 
        
    \item Laplacian Matrix: Given a simple graph $G$, the \textit{Laplacian Matrix} of $G$ is defined as $L = D -A$, where $D$ is the diagonal degree matrix and $A$ is the adjacency matrix of $G$. If a graph $G$ is undirected, its \textit{Laplacian Matrix} is also a symmetric matrix. It can be normalized to matrix of unit vectors, usually denoted as $\curlyL = D^{-1/2}LD^{1/2} = I - D^{-1/2}AD^{1/2}$. 
\end{itemize}

\section{Network Motifs}
\label{sec:motif}
Compared to more complex higher-order network representations, motifs have been studied for a relatively longer period. There are two main reasons: (1) motifs are studied directly on dyadic graphs, which are widely used in various research fields; (2) some specific subgraphs already have some special meanings in the real world, so it is natural to study them in networks.

\subsection{Foundation and Algorithms}
A network motif is generally defined as \textbf{a highly significant subpattern or subgraph in the network}~\cite{alon-network-motifs}. The term ``significant'' indicates that the number of times the motif appears in the graph is higher than what is expected or normal, where such expected numbers are often understood within \textit{random graphs} (e.g., the \textit{Erdős–Rényi model}~\cite{erdos59a}). A motif can be some fixed-size subgraph such as a triangle; or can have a variable size representing some general conceptual pattern such as a star or a loop~\cite{ShenOrr2002NetworkMI}. For brevity, we focus on fixed-size subgraphs. Formally, a network motif is
\begin{definition}[Network motif]
\label{def1:net_motif}
Motif $M$ of graph $G$ is a subgraph of $G$ that has multiple isomorphic graphs that are also subgraphs of $G$. That is, $G_1 \subseteq G, G_2 \subseteq G, \dots, G_n \subseteq G$ and $G_1,...,G_n$ are isomorphic to $M$.
\end{definition}

\subsubsection{Motif Frequency}
For motif $M$, its number of appearances in graph $G$ can be denoted as $F_G(M)$. By comparing this frequency with the mean (expected) count in random graphs with the same size as $G$, we obtain the $Z$-score of motif frequency 
\begin{equation}
\label{eq:zscores}
    Z(M) = \frac{F_G(M)-\mu_R(M)}{\sigma_R(M)},
\end{equation}
where $\mu_R(M),\sigma_R(M)$ are the expected mean and standard deviation of frequencies of $M$ in a random graph. This $Z$-score is an important statistic for measuring the significance of motifs. 

\subsubsection{Motif Matrix and Motif Cuts}
Similar to the adjacency matrix, Benson et al.~\cite{doi:10.1126/science.aad9029} define \textit{motif adjacency matrix} based on the memberships of edges in the given motifs. Formally, given a specific motif $M$, the motif adjacency matrix $W_M$ is defined as 
\begin{equation}
    (W_M)_{ij} = \Big|\{M|i \in M , j\in M\}\Big|,
\end{equation}
where $(W_M)_{ij}$ is the number of instances of $M$ that contain nodes $i$ and $j$. When the given motif is an edge (two connected nodes), the motif adjacency matrix is simply the adjacency matrix. Note that when the given motif is not a complete graph, the nodes $i$ and $j$ can belong to the same motif even if they are not connected. 

The motif adjacency matrix is also of size $|V|\times|V|$, so most algorithms designed for the adjacency matrix are also suitable for the motif adjacency matrix. 

With the motif adjacency matrix in place, a \textit{motif cut} can be  defined consequently for
motif $M$ and motif adjacency matrix $W_M$: 
\begin{equation}
    cut_M(S,\Bar{S}) =\sum_{i\in S,j\in \Bar{S}}W_{Mij}. 
\end{equation}

Similarly, \textit{motif conductance} is defined as 
\begin{equation}
    \varphi_M(S) = \frac{cut_M(S,\Bar{S})}{min(\text{Vol}_M(S),\text{Vol}_M(\Bar{S}))},
\end{equation}
where $\text{Vol}_M(S)$ denotes the volume (total sum of edge weights) of set $S$ in the motif matrix. If a clustering algorithm minimizes motif conductance, the result leads to preserving the structure of the given motif in the graph while generating a balanced split by the clustering algorithm. 

\subsubsection{Motif Clustering Coefficient}
\label{sec:higherCC}
The \textit{clustering coefficient} is a measure of the degree to which nodes in a graph tend to cluster together~\cite{watts_collective_1998}. In dyadic graphs, the clustering coefficient can be calculated by the fraction of length-2 paths (wedges) that are involved in triangles. From this perspective, the \textit{global clustering coefficient} can be defined as  
\begin{equation}
    C = \frac{6|K_3|}{|W|},
\end{equation}
where $|K_3|$ is the number of triangles (3-cliques), and $|W|$ is the number of wedges. Each triangle is counted six times since it contains six different wedges, considering the order. The \textit{local clustering coefficient} of node $u$ is defined as 
\begin{equation}
    C(u) = \frac{2|K_3(u)|}{|W(u)|},
\end{equation}
where $C(u)$ is the fraction of triangles that node $u$ belongs to over the number of wedges in which $u$ is the center node. 

Based on the above interpretation of the clustering coefficients, Yin et al.~\cite{2018honhao} generalize higher-order clustering coefficient to motifs of higher-order cliques. For order $l\geq 2$, clustering coefficient of order $l$ can be defined as 
\begin{equation}
    C_l = \frac{(l^2+l)|K_{l+1}|}{|W_l|},
\end{equation}
where $K$ and $W$ are higher-order cliques and wedges, and $(l^2+l)$ is basically $(l+1)l$, which is the number of wedges that an $(l+1)$-clique closes. Consequently, the local higher-order clustering coefficient is generalized as 
\begin{equation}
    C_l(u) = \frac{l|K_{l+1}(u)|}{|W_l(u)|}.
\end{equation}

\subsection{Use Cases and Applications}
Motifs are utilized across many scientific fields. Here, we survey use cases and applications of motifs.

\subsubsection{Capturing Functionalities in Networks} 
Highly frequent motifs are often related to specific functionalities that networks capture, especially in biology. We provide some examples in biological and brain networks.

\vspace{1mm}\textbf{Biological networks.} In 2002, Shen-Orr et al.~\cite{ShenOrr2002NetworkMI} distinguished three families of motifs in Escherichia coli (E. coli) directed transcriptional network. These families are closely related to specific functionalities. They are: feedforward loop (a directed acyclic graph), single input module (one to many transactions) and dense overlapping regulons (many to many transactions). Each motif relates to a specific function in the determination of gene expression. Such frequent motifs can be detected using a brute-force approach on some sub-matrix of the adjacency matrix. 

The functionality of motifs in biological networks is further validated through a simulation of \textit{spontaneous evolution process}~\cite{Kashtan13773}. First, an electronic combinatorial logic circuit is initiated by random wiring. The goal of network evolution is to increase the fraction of correct output under given logical functions. The baseline, a \textit{fixed goal} given by $G_1=(X \oplus Y) AND (Z \oplus W)$, is very slow to converge with a low rate of reaching the perfect solution. Networks that evolved under fixed goal have less significant motifs and lower modularity (the separability of the design into units that perform independently). Addressing this issue, the authors introduce \textit{modularly varying goals}, where the goals switch between $G_1$ and $G_2=(X \oplus Y) OR (Z \oplus W)$ every 20 epochs. Surprisingly, such evolved networks could always find perfect solutions of the current goal (either $G_1$ and $G_2$) quickly within a few epochs. Similar findings are discovered in neural networks, which explain adaptiveness and robustness of motifs in real-world biological networks.


\vspace{1mm}\textbf{Brain Networks.} Due to the complexity of brain networks, relationships between subareas of the brain and their functionality are of significant research interest. In brain networks, motifs are often divided into two groups: \textit{functional} and \textit{structural}. Structural motifs are those currently presented in this survey and capture the anatomical building blocks of the brain network, whereas functional motifs capture patterns of elementary processing within such structural motifs. In other words, functional motifs can be considered as all possible subgraphs of the structural motif with the same number of nodes but different edges. For example, a triangle structural motif consists of three different functional motifs (all paths of length two). One hypothesis suggests that the number and variety of \textit{functional motifs} are maximized in the brain to increase effectiveness~\cite{brain_article}. 

Functional motifs vary significantly over time. Duclos et al.~\cite{Duclos2020BrainNM} investigate motif appearances in the brain network over time. They count all connected subgraphs of size three in directed brain networks, and as a result, show that anesthetic-induced unconsciousness is associated with a topological re-organization of the brain network. Specifically, the frequency of chain-like and loop-like motifs change significantly when people transition from a responsive state to an unresponsive state. Such observations demonstrate links between motifs and functionalities.

In terms of the shape of motifs in brain networks, research has been more interested in specific motifs that are easier to count, such as cliques and \textit{cavities} (enclosed spaces). For instance, all maximal cliques are counted, and their frequency is compared to that of what is expected in some null model. The results indicate the spatial distributions of maximal cliques are more than expected in different brain regions. Similarly, cavities can be studied (ranging from minimum cycles to incomplete cliques). Research shows that, unlike cliques, cavities are less than expected in different areas of the brain~\cite{articleclique}. 

\subsubsection{Network Classification} 
In the seminal work of Milo et al.~\cite{alon-network-motifs}, network motifs are defined as specific fixed subgraphs whose frequencies are higher than what one would expect in random graphs. It turns out motifs found and their frequencies from various types of networks (including food webs, biological networks, electronic circuits, World Wide Web, and the like) can be utilized to classify types of networks. In particular, graphs from the same category exhibit significant overlap in terms of motifs observed and their frequencies, which can be used as features to distinguish graphs. For example, on food webs, a three-node chain is frequently observed. However, this motif is not frequently observed in any other category of networks. The feed-forward loop is popular on most information-processing networks, including brain networks, electronic circuits, and the World Wide Web. 

Milo et al. further investigate the number of motifs across various categories~\cite{doi:10.1126/science.1089167}. $Z$-scores (see Equation \ref{eq:zscores}) are calculated for subgraph frequencies and are compared with those expected in random graphs. The results show that the graphs from the same category have highly similar subgraph frequencies. However, some graphs from different categories also show similarities in their frequencies, which captures the intrinsic similarity between graphs that are similar categories. As a result of this discovery, networks from different categories can be classified into superfamilies using motif frequencies. 

\subsubsection{Network Models}

As motif frequencies are different in real-world graphs compared to what one would expect in random graphs, there is a need for \textit{network models} that can generate realistic random graphs with motif frequencies similar to those of real-world graphs. 

Pržulj et al.~\cite{10.1093/bioinformatics/bth436} propose a new series of network models called geometric random graphs, which uniformly generate nodes (i.e., points) in 2D/3D/4D Euclidean space and form links between nodes based on a threshold on their distances. By counting all possible motifs under size five, they found that the random graphs generated by geometric models are more similar in motif counts to the original protein-protein networks than those generated by other network models.

There is a significant need to develop network models that can generate motifs with similar frequencies to those observed in real-world data.
One example is the network model designed by Leskovec et al.~\cite{leskovec2010signed}. The work examines all triangles in \textit{signed} networks, where edges can have a sign: $+$ or $-$. For instance, a $+$ edge may indicate that two nodes are ``friends," and a $-$ edge may indicate that two nodes are ``enemies." They discovered
that network models that simulate the \textit{balance theory} (colloquially stated as ``an enemy of an enemy is my friend") might be insufficient to explain the frequency of triangles appearing in real-world networks. Therefore, they propose another network model for directed links, inspired by \textit{status theory}, where a positive link from node $a$ to $b$ indicates that $a$ has a higher ``status" than $b$. Given a three-node directed cycle with two positive signs, these two theories will predict opposite signs for the third link. Balance theory explains this as three pairwise friends (friend of a friend is a friend), while status theory considers $a\rightarrow b\rightarrow c$ as a pattern of increasing status, so $c$ should link to $a$ with a negative sign (high to low status). In directed graphs, research shows that graphs generated based on status theory could more realistically replicate the frequencies of signed motifs compared to graphs generated based on balance theory. 

\subsubsection{Clustering}
Motifs are closely related to clusters in higher-order networks. Benson et al.~\cite{doi:10.1126/science.aad9029} develop a framework of network clustering based on higher-order properties. More specifically, cuts on edges are generalized to cuts on motifs, and the adjacency matrix is generalized to the motif membership matrix. Their results show that such clustering accurately preserves higher-order structures. 
Two real-world examples are presented. For clustering based on a \textit{bi-fan} motif, the clustering clearly distinguishes between the role of source and sink by assigning them to different clusters in neuronal networks. In the airline network, transportation hubs are clustered together by using a bi-directional 2-path motif. 

Building upon the ideas of motif matrix and motif cuts, Yin et al.~\cite{10.1145/3097983.3098069} generalize clustering methods to the motif level. They first propose a motif-based approximate personalized PageRank (MAPPR), which performs an approximation of the Personalized PageRank using the motif matrix. The method can quickly find a cluster that contains a given seed node that has the minimum \textit{motif conductance}. To enhance the performance in case the clustering is performed on the whole graph, they introduce an efficient method to identify good seed nodes to be used as input to MAPPR. 
The proposed method is validated by performing cluster recovery tasks on synthetic and real-world graphs. Experimental results show that the proposed techniques could preserve higher-order clustering coefficient (as detailed in Section \ref{sec:higherCC}). 

Furthermore, as we also showed in Section \ref{sec:higherCC}, some traditional measurements of clusters (or clusterability) are generalized to the motif-level. One important generalized graph measurement was the \textit{clustering coefficient}, which reflects the degree of cohesiveness as communities. Yin et al.~\cite{2018honhao} introduced higher-order variants of the local and global clustering coefficients. For order-3, the global clustering coefficient is defined as the ratio of cliques to wedges (length-2 paths); the local clustering coefficient is defined as the ratio of cliques that a node involved over wedges. Interested readers can refer to Section \ref{sec:higherCC} for extensions of clustering coefficients to orders greater than three. 

Another important measure for a clustering is its \textit{modularity}. Modularity is a quantitative measure to evaluate the significance of clusters, which is also generalized to the motif level~\cite{10.1093/comnet/cnu016}, specifically two special motifs -- cycles and paths. In the general case, motif modularity is defined as the fraction of motifs laying fully inside the community subtracted by that of expected in the random graphs. Higher-order modularity can distinguish differences in higher-order structures, such as cycles and cliques/hubs and leaves. For example, in a multipartite network roles can be easily distinguished simply by applying higher-order modularity. 

\subsubsection{Representation Learning}
Representation Learning aims at encoding specific network properties into fixed-length vectors. 
In order to capture higher-order structures, Rossi et al.~\cite{rossi2018hone} propose a network embedding method based on motifs. First, they build several weighted motif adjacency matrices based on the nodes' occurrences in specific motifs. Then they define a series of functions over these weighted motif matrices, such as $k$-step paths, the transition matrix, and various Laplacians. By minimizing the distances between the motif-based matrix formulation and the embedding matrix, each motif matrix learns a local embedding. Finally, they concatenate the local embedding to calculate a global embedding for the network. Experimental results show that the proposed higher-order network embeddings outperform other embedding methods in link prediction tasks.

Another higher-order representation learning method using motifs is \textit{LEMON}~\cite{10.1145/3473911}. First, \textit{LEMON} converts the graph by adding \textit{supervertices} for motifs (e.g., triangles), and then links the nodes that are involved in such motifs to the corresponding supervertices. The result is a \textit{two-mode} network that captures the memberships of nodes in motifs, where the edges between motif supervertices to regular nodes are defined as structural edges. The embedding vector is learned through a random walk process that captures the similarities of nodes that share similar motif structures. A parameter $q$ controls the traversal probability from a regular node to supervertices. With larger $q$, any node will become closer to nodes similar in motif properties rather than its structural neighbors. \textit{LEMOM} has been successfully applied in anomaly detection, link prediction, and node classification.

\subsubsection{Link Prediction}

Motif counts can be used a powerful feature to predict missing links.
Abuoda et al.~\cite{DBLP:journals/corr/abs-1902-06679} convert the link prediction problem into a classification problem by counting the motifs involved in the link candidates. All possible connecting motifs within size five are enumerated as features. The classification results of several classical classifiers show that larger motifs can lead to higher performance and that a combination of motifs can further improve the results. The work shows that motif-combined feature classification outperforms most state-of-the-art link prediction methods.

\section{Simplicial Complexes}
\label{sec:simplicial_complexes}
A simplicial complex can be interpreted as another generalization of a graph. In graphs, there are two different types of entities -- nodes and edges. But in simplicial complexes, the concepts of nodes and edges are merged into a generalized basic unit, the \textit{simplex}, where $0$-simplex represents the single vertex and $1$-simplex represents the edge. Furthermore, a simplicial complex can contain any \textit{order} of interactions, such as $k$-simplices ($k\geq0$). 

The main difference between simplicial complexes and hypergraphs is the requirement of being \textit{inclusive}; that is, a simplicial complex also contains all \textit{faces} (sub-simplices) of its current simplices. For example, if three people $A$, $B$, and $C$ belong to the same university, then all pairs: $AB$, $AC$, and $BC$ have the same relationship (being part of the university). When modeling higher-order data with simplicial complexes, ensuring the relationship being modeled is inclusive is the first requirement. 

\subsection{Foundation and Algorithms}
In mathematics, a simplicial complex is a set composed of points, line segments, triangles, and their $n$-dimensional counterparts. 

\subsubsection{Simplex}
A \textit{simplex} is the basic unit of a simplicial complex and is the generalization of the notion of a triangle or a tetrahedron to higher dimensions. More specifically, a \textit{$k$-simplex} is a $k$-dimensional polytope that is the convex hull of its $k + 1$ vertices. 

The convex hull of any nonempty subset of the $k + 1$ points that define a $k$-simplex is called a \textit{face} of the simplex. Faces are also simplices. Any $k-1$-face of a $k$-simplex is called a \textit{facet}.

\subsubsection{Simplicial Complex}
\begin{definition}[Simplicial Complex]
\label{def2:simplicial complex}
A simplicial complex $\mathcal{X}$ is a \underline{set} of simplices that satisfies the following:
\begin{enumerate}
    \item Every face of a simplex from $\mathcal{X}$ is also in $\mathcal{X}$; and
    \item The non-empty intersection of any two simplices $\sigma_1,\sigma_2 \in \mathcal{X}$ is a face of both $\sigma_1$ and $\sigma_2$.
\end{enumerate}
\end{definition}
Roughly speaking, simplicial complexes are simplices that are (1) closed under taking faces and (2) have no inner intersections other than faces. A \textit{simplicial $k$-complex} $\mathcal{X}$ is a simplicial complex where the largest dimension of any simplex in $\mathcal{X}$ equals $k$. 

\subsubsection{Homology}
First, we define the \textit{orientation} of a simplex. The orientation of a $k$-simplex is given by an ordering of the vertices $(v_0, v_1,\dots,v_k)$. There are exactly two orientations -- even and odd permutations, and switching any two vertices in the ordering leads to a change of the orientation. For example, in a two-dimensional space, we have clockwise and counterclockwise ordering for a triangle. The orders $(v_1,v_2,v_3)$,$(v_2,v_3,v_1)$,$(v_3,v_1,v_2)$ indicate one orientation, and the orders $(v_1,v_3,v_2)$,$(v_3,v_2,v_1)$,$(v_2,v_1,v_3)$ indicate the opposite one. 

Let $\mathcal{X}$ be a simplicial complex. A \textit{simplicial $k$-chain} is a finite formal sum
\begin{equation}
    \sum_{i=1}^{N} c_i\sigma_i,
\end{equation}
where $c_i$ is an integer and $\sigma_i$ is an oriented simplex. For each simplex, the sum includes a sign based on the orientation. One way of assigning orientations is to order all vertices of the simplicial complex and give each simplex the orientation corresponding to it. The group of $k$-chains on $\mathcal{X}$ is written $C_k(\mathcal{X})$, and for simplicity we write $C_k$. Note that $C_k$ is a vector space with the number of $k$-simplices as its dimension. 

Based on the group of $k$-chains $C_k$, we define \textit{boundaries} and \textit{cycles}. First, we define the boundary operator. 

\begin{definition}[Boundary Operator]
\label{def3:boundary operator}
Let $\sigma = (v_0,...,v_k)$ be an oriented $k$-simplex, viewed as a basis element of $C_k$. The boundary operator $\partial_k : C_k \rightarrow C_{k-1}$ is the homomorphism defined by:
$$\partial_k(\sigma) = \sum_{i=0}^{k}(-1)^i(v_0,\dots,\widehat{v_i},\dots,v_k),$$
\end{definition}
\noindent where $(v_0,\dots,\widehat{v_i},\dots,v_k)$ is the $i^{th}$ face of $\sigma$, which deletes $v_i$ from $\sigma$. 

The boundary of each $k$-simplex is the collection of all its $(k-1)$-faces. In $C_k$, elements of the subgroup $Z_K:=\ker \partial_k$ are referred to as \textit{cycles}, which is the collection of $k$-simplexes whose boundary is zero. While subgroup $B_K:=\Ima \partial_{k+1}$ denotes the boundaries, i.e., boundaries of $(k+1)$-simplices. Note that using definition~\ref{def3:boundary operator}, it is easy to prove that the boundary of boundaries is empty. 

Cycles are essential entities for detecting holes. However, some simplices in $Z_K$ are just boundaries of $(k+1)$-simplices, which are not holes themselves. Hence, we remove the boundaries of the $(k+1)$-simplices from the cycles. This can be defined as the \textit{quotient abelian group}
\begin{equation}
    H_k = Z_k/B_k = \ker \partial_k/\Ima \partial_{k+1},
\end{equation}
where the remaining simplices in $H_k$ represent $k$-dimensional holes in the complex. $H_k$ is called the \textit{homology group}. 

\subsubsection{Cohomology and Hodge Laplacian}
Remember the group of $k$-chain $C_k$ is a vector space over $\mathbb{R}$. Hence, it is possible to give an inner product structure to each $C_k$ to make the basis (oriented simplices) orthogonal. We denote this dual space of $C_k$ as $C^k$~\cite{Muhammad06controlusing}, called the group of \textit{$k$-cochains}. 

We denote the dual operator of the boundary map $\partial_k$ as $\delta_k$. Operator $\delta_k: C^{k} \rightarrow C^{k+1}$ is called the \textit{co-boundary} operator, which is the adjoint of boundary map $\partial_k$. Consequently, the \textit{cohomology} group is defined over \textit{cochains} 

\begin{equation}
    H^k = \ker \delta_k/ \Ima \delta_{k-1},
\end{equation}
which is exactly a dual group of the homology group $H_k$. Note that the cohomology groups are defined more algebraically with less geometric meaning. The main purpose of introducing the cohomology group is to derive the \textit{Hodge Laplacian} (see Definition \ref{def5:HodgeLaplacian}). For a detailed explanation, interested readers can refer to~\cite{DBLP:journals/corr/Lim15}. 

Given a simplicial complex $\mathcal{X}$, its boundary map $\partial_k$ can be represented as a matrix $B_k$. $B_k$ has the dimension $n_{k-1}\times n_{k}$, where $n_{k-1}$ and $n_{k}$ are the number of $(k-1)$-simplices and $k$-simplices, respectively. For example, $B_0=0$ and $B_1$ is a matrix of dimension $|V|\times |E|$. 

Similarly, the co-boundary map $\delta_k$ can also be represented as the adjoint matrix $B_k^*$. In a finite real space, it is equal to the transpose of $B_k$, so we can also write it as $B_k^\intercal$.

\begin{definition}[Hodge Laplacian]
\label{def5:HodgeLaplacian}
The \textit{$k$th Hodge Laplacian} of a simplicial complex $\mathcal{X}$ is defined as 
\begin{equation}
    \mathcal{L}_k = B_k^\intercal B_k+ B_{k+1}B_{k+1}^\intercal.
\end{equation}
\end{definition}
When $k=0$, $\mathcal{L}_0 = B_{1}B_{1}^\intercal$ is exactly the Laplacian matrix in dyadic graphs, with dimension $|V|\times |V|$. Matrix $\mathcal{L}_1$ has dimension $|E|\times |E|$, capturing relationships among basic units of edges~\cite{DBLP:journals/corr/abs-1807-05044}. 

\subsubsection{Degrees and Random Simplicial Complex}
Degree is generalized in simplicial complex as follows~\cite{Courtney_2016}:
\begin{definition}[Degree of a Simplex]
\label{def5:degree}
For any simplex $\sigma \in \mathcal{X}$, the degree $k_{d,\lambda}(\sigma)$ is the number of $d$-dimensional simplices adjacent with $\sigma$ in $\lambda$-faces. 
\end{definition}
When we are only interested in the degree of vertices, we let $\lambda = 0$. Then $k_{d}(v)$ becomes the number of $d$-simplex incident to $v$, i.e., those that $v$ belongs to. 

\vspace{1mm}As an analog to the \textit{Erdős–Rényi} model~\cite{erdos59a} in dyadic graphs, the generative model of 2-complexes can be defined as follows: 
\begin{definition}[Random 2-Complex]
\label{def4:2complex}
The $\mathcal{X}(n,p)$ model of a simplicial complex is defined to have vertex set $[n]$, edge set $[n]\choose 2$, and each of the $n\choose 3$ possible triangle faces is included independently with probability $p$.
\end{definition}
Note that for a 2-complex both nodes and edges (e.g. a complete graph) have to be specified, and the random process only occurs on random triangles~\cite{https://doi.org/10.48550/arxiv.1301.7165}. One can define random simplicial complexes of higher order in similar ways. 

\subsection{Use Cases and Applications}
Applications of the simplicial complex have mainly focused on two directions. One direction is focused on topological properties, where a simplicial complex is used to represent a space. In many fields, the real-world information can be abstracted to pure topology entities through a process called \textit{filtration}, which transforms real distances into topological edges. Such techniques are widely used in sensor coverage problems, biological networks, mobility analysis, robotics, and the like. The second direction is to model real-world interactions of more than three individuals as simplices. 

\subsubsection{Sensor Coverage}
Sensor coverage problem aims at measuring a ``coverage" area and detecting locations that are uncovered: also known as \textit{holes}. 
Ghrist and Muhammad~\cite{Ghrist2005Coverage} modeled the sensor coverage problem using simplicial complexes. A coordinate-free sensor network is formed by relative distances between any pair of nodes, without any specific coordinates. This is simpler to obtain through the strength of signals sent by the sensors, especially in dynamic systems. Based on simplicial homology theory, coverage holes are what remain after removing boundaries from cycles. The theoretical results are also verified by practical simulations in computational homology software.

The sensor cover can be further linked to the homology of the diagram of complexes~\cite{De_coveragein}. In particular, the sensor cover can be defined as a collection of discs of radius $r_c$, and the radius of strong and weak signals of pairwise distances can be represented as $r_s$ and $r_w$. \textit{Rips complex} is defined as a simplicial complex whose simplices are tuples of nodes whose pairwise Euclidean distances are within a certain threshold. Each node can detect the existence of the boundary of the domain within another radius $r_f$. By forming the simplicial complexes of all these graphs, one can derive the sensor coverage. 

Under similar settings with previous studies, Tahbaz-Salehi and Jadbabaie~\cite{5444998} present a distributed algorithm for coverage verification without any metric information. The goal of coverage verification is basically three aspects -- detecting coverage holes, calculating their locations, and detecting redundancies in the network. The main novel contribution of this work is to solve the homology problem through a linear programming relaxation. 

\subsubsection{Disease/Abnormality Detection}

Point cloud is one of the classic data formats often used in biology, where points are substances such as proteins. Similar to the sensor coverage problem, a simplicial complex can be constructed over a point cloud through the \textit{filtration} process~\cite{Nanda2014SimplicialMA}. Points agglomerate together and become simplicies when their distances fall under a specific threshold, specified by some distance function. One notable usage is to identify subtypes of breast cancer. For example, a simplicial complex can be used for preprocessing to enhance the clustering performance~\cite{10.1093/bioinformatics/btm033}. Simplicial complexes can also help distinguish between recurrent and non-recurrent subtypes~\cite{DeWoskin2010ApplicationsOC, DBLP:journals/aaecc/ArsuagaBDMPP12}.

In brain networks, a new topology called \textit{homological scaffold} can be defined to represent low-connection areas in the network~\cite{doi:10.1098/rsif.2014.0873}. First, a brain network can be seen as a weighted network, where larger weights indicate longer distances. Then a filtration process is applied to generate sparse structures (larger weights), followed by the detection of a homology group. The remaining structure in the homology group contains cycles with larger distances, which captures areas in the network that exhibit extremely low connections. 

In neuroscience, for amplifying the differences in network topologies, a novel matrix signature is proposed to facilitate forming the homology groups~\cite{doi:10.1073/pnas.1506407112, https://doi.org/10.48550/arxiv.1601.01704}. Instead of the absolute distances, the orders of distance are used. For example, if the distance of $v_0$ and $v_1$ is the minimum of all pairwise distances, then the entry of $A_{01}$ will be encoded as 0. Such a non-linear transformation obscures the distances but focuses more on the intrinsic structure of the network. The order matrix captures a more robust relationship with the topological structure, for example, the number of noncontractible cycles. Experiments on pyramidal neurons in the rat hippocampus show that the proposed signature is capable of detecting geometric organization.

\subsubsection{Mobility Analysis}

To study mobility, topological signatures have been proposed that represent trajectories as points in $k$-dimensional space, where $k$ is the number of obstacles~\cite{10.1145/3274895.3274952}. The goal of this mapping is to characterize the differences in traces when passing by obstacles. First, obstacles are represented as simplicial complexes, and any motion toward faces can be recorded by sensors. Based on homology, these faces (edges) are encoded as real numbers. These values are added to the entries of the related obstacles as trajectories records. Then, trajectory traces can be distinguished by this signature. For example, one coming across an obstacle from the left is encoded as 1, while as -1 if coming from the right; if one loops clockwise around an obstacle, we can encode that as 2, and -2 for counterclockwise loops. 

\subsubsection{Network Modeling}

The configuration model is a method for generating random networks from a given degree sequence. For a simplicial complex, the configuration model is also generalized along with the \textit{canonical ensemble}~\cite{Courtney_2016}. In short, the canonical ensemble aims to derive the probability of the simplicial complex that maximizes the entropy defined by it. The configuration model is the uniform distribution of all possible simplicial complexes with the same degree sequence. 

Based on such generalizations, Young et al.~\cite{Young_2017} further develop efficient sampling algorithms for the simplicial complex configuration model. First, they elaborate the numerical constraint of the configuration model by switching the simplicial complex to its equivalent \textit{graphical} representation. That is, to introduce extra nodes to represent adjacent relationships between nodes and simplicies. In this way, the simplicial complex is transformed into a dyadic bipartite graph, which can yield a solution~\cite{https://doi.org/10.48550/arxiv.1004.2612}. 

The \textit{social contagion} can be modeled as a propagation network, where people get infected through social interactions (edges). In terms of a simplicial complex, a contagion model could also consider an infection being caused by a group, called \textit{Simplicial Contagion Model}~\cite{2019simplicialm}. Similar to the dyadic contagion model, any pairwise interaction can lead to an infection with a uniform probability ($\beta_1$). In addition, higher-order interactions have unique contagion probabilities if there are multiple infecteds involved. For example, in a simplicial complex, if the other two nodes are infected, the candidate will have a probability of ($\beta_\Delta$) being infected. The behavior of the infection pattern is discussed by simulating the process over both real-world and synthetic graphs. The \textit{Simplicial Contagion Model} is a more flexible fit for more complex diseases with varying infection probabilities of different orders. 

In quantum physics, Bianconi and Rahmede~\cite{Bianconi_2017} propose a model of emergent geometry that is based on a growing simplicial complex. The model is simple as it just keeps including simplicies with fixed dimension $d$ and gluing them to the existing simplicial complex on one of its $d-1$ faces. Many advantages of such a model are validated and discussed under certain settings, including scale-free degree distribution, small-world properties, and modular structure. 

\subsubsection{Network Analysis Tools}

Instead of studying motifs in dyadic networks, Benson et al.~\cite{DBLP:journals/corr/abs-1802-06916} directly collect higher-order relationships in the real world, such as co-authorships, event participation, drug instances, among other similar interactions. Such coappearances are modeled as simplicial closures (timestamped vertex sets). For example, a closed triangle indicates relationships among three nodes, while an open triangle just represents pairwise relationships between any two nodes. This representation enriches the network information and can be used in dynamic graph-evolving models or link predictions. In the link prediction task, the goal is to predict whether the open triangles will become close in the future. Results show that even simple local features such as the mean of weights on three edges perform pretty well and are comparable with state-of-the-art methods. 

Based on the $1$st normalized Hodge Laplacian, Schaub et al.~\cite{DBLP:journals/corr/abs-1807-05044} discuss random walks on basic units of edges in a simplicial complex. This work enriches the traditional field of network analysis, which is mostly node-based. Two applications are performed to verify the usage. One is representation learning of edge-flows and trajectory data, as a higher-order generalization of diffusion maps and Laplacian eigenmaps. Another is the edge-based generalization of PageRank~\cite{DBLP:journals/corr/Gleich14}, which focuses on the importance of edges rather than nodes. 

Regarding clustering, Osting et al.~\cite{DBLP:journals/corr/abs-1708-08436} applied a sparsification process on a simplicial complex, which downgrades the maximum dimension under a given threshold. It is proved that such a sparsification preserves the \textit{up Laplacian}. The authors also generalize \textit{Cheeger inequalities} to a simplicial complex. The preservation of the spectrum is verified through experiments, and spectral clustering is performed as one application. 

Advanced network representation learning methods have also been extended to the simplicial complex. Hajij et al.~\cite{DBLP:journals/corr/abs-2103-04046} use the autoencoder to perform simplex-level embedding. The encode function ($X\rightarrow \mathbb{R}^d$) maps each simplex to a fixed vector. The decode function ($\mathbb{R}^d \times \mathbb{R}^d \rightarrow \mathbb{R}^+$) maps each pair of simplices to a similarity score that reflects the relationship between the two simplices. An example of a user-defined similarity is the simplex-level adjacency matrix. Finally, the representation of the whole simplicial complex is obtained by a weighted sum of the simplex-level representations. 

\section{Hypergraphs}
\label{sec:hypergraph}
Another natural generalization of graphs is called hypergraphs, which extend the edge space from $|V|^2$ to $|V|^{|V|}$. In other words, an edge of a hypergraph can be any subset of the vertices. The main difference with the simplicial complex is that any subset of an edge can exist independently from others. 

Though easy to understand, the extremely sparse and high-dimensional edge space causes difficulty in computations. So, relatively more studies on hypergraphs have focused on the theoretical aspects rather than applications. For a more comprehensive review of concepts and measurements in hypergraphs, we refer interested readers to the survey by Lee et al.~\cite{lee2024survey}.

\subsection{Foundation and Algorithms}
Many definitions and properties of hypergraphs are directly inherited from graphs, such as \textit{node degrees}, \textit{hyperedge weights}, \textit{simple/multi} hypergraphs, hypergraph \textit{isomorphism}, and so on. Here, for brevity, we mainly focus on the definitions unique to hypergraphs. 

\subsubsection{Definition and Basics}
We use $\mathcal{H} = (V, \mathcal{E})$ to distinguish a hypergraph from a graph, where only vertex set $V$ remains the same. The edge set $\mathcal{E}=\{e|e\subseteq V \}$ is the set of subsets of $V$. We define \textit{size} of edge $|e|$ as the number of nodes that belong to edge $e$. 

We can have special kinds of hypergraphs:
\begin{itemize}
    \item $d$-regular: each vertex has degree $d$;
    \item $k$-uniform: each edge has size $k$;
    \item $k$-partite: vertices belong to one of $k$ different classes, and each edge has exactly one node from each class.  
\end{itemize}

A hypergraph can be always represented by a bipartite graph of vertices and edges. The \textit{biadjacency matrix} of this bipartite graph is a $|V|\times |E|$ matrix, which is also called the \textit{incidence matrix} of the hypergraph. 

\subsubsection{Tensor Representation}
The natural generalization of the adjacency matrix for the hypergraph is a \textit{tensor}, often denoted by $\mathbf{T}$. For example, a $3$-uniform hypergraph (or the subset of order-3 edges) can be represented as an order-3 tensor $\mathbf{T} \in \mathbb{R}^{|V|\times|V|\times|V|}$, i.e., the entry $(i,j,k)=1$ when $(v_i,v_j,v_k)\in \mathcal{E}$. If the hypergraph is undirected, the corresponding tensor is \textit{symmetric}, where its value at any permutation of $(i,j,k)$ remains the same. A \textit{simple tensor} can be written as the outer product of the vectors. The \textit{rank} of a tensor is the minimum number of simple tensors whose linear combination equals that tensor. 

\vspace{1mm}\textbf{(Tensor Decomposition)} Due to tensor dimensionality, it is expensive and inconvenient to perform calculations directly on it. Decomposition techniques are widely used to decrease the dimension and preserve the graph characteristics. Here, we introduce two popular decomposition methods -- CP decomposition and Tucker decomposition. For a detailed reference on the decomposition techniques of the tensor, interested readers can refer to~\cite{7891546}.

CP (CANDECOMP/PARAFAC) Decomposition~\cite{Carroll1970AnalysisOI, Harshman1970FoundationsOT} is also called \textit{tensor rank decomposition} or \textit{Canonical Polyadic Decomposition (CPD)}. The CP decomposition factorizes a tensor into a sum of component vectors. For example, a tensor $\mathbf{T} \in \mathbb{R}^{I\times J\times K}$ can be decomposed as 
\begin{equation}
    \mathbf{T}\approx \sum_{i=1}^{R}\mathbf{a_i}\otimes \mathbf{b_i}\otimes \mathbf{c_i},
\end{equation}
where $R$ is an integer and $\mathbf{a_i}\in \mathbb{R}^I$, $\mathbf{b_i}\in \mathbb{R}^J$, $\mathbf{c_i}\in \mathbb{R}^K$, and $\otimes$ denotes the outer product sign. This decomposition is often solved by some minimization algorithm. 

Tucker Decomposition~\cite{Tuck1966c} is the generalization of \textit{Singular Value Decomposition (SVD)} for the tensors. Tucker decomposition of a tensor $\mathbf{T} \in \mathbb{R}^{I\times J\times K}$ is represented as 
\begin{equation}
    \mathbf{T}\approx \mathcal{G} \times \mathbf{A}\times \mathbf{B}\times \mathbf{C}=[\![\mathcal{G};A,B,C\mathbb]\!],
\end{equation}
where $\mathbf{A}\in \mathbb{R}^{I\times P}$, $\mathbf{B}\in \mathbb{R}^{J\times Q}$, $\mathbf{C}\in \mathbb{R}^{K\times R}$. Here, tensor $\mathcal{G}\in \mathbb{R}^{P\times Q\times R}$ is called the \textit{core tensor}. 

\subsubsection{Hypergraph Cuts}
In dyadic graphs, a cut is defined as partitioning the vertices into two disjoint subsets, where the cut edge connects two nodes, one from each subset. However, for a hyperedge, there is no such fair split where we cannot assign some nodes to one side and the rest to the other side. Among various cut functions, there is only one that might be more reasonable to minimize when solving cut-based hypergraph problems, called \textit{all-or-nothing}~\cite{DBLP:journals/corr/abs-2001-02817}. 

Specifically, a hyperedge is in the middle of the cut when it is assigned to both sides of the cut. One can present the set of \textit{cut hyperedges} by
\begin{equation}
    \mathbf{\partial}S = \{e\in \mathcal{E}:e \cap S \neq \emptyset \;and\; e \cap \Bar{S} \neq \emptyset\}.
\end{equation}
The cut function is
\begin{equation}
    \textbf{all-or-nothing}(s) = \sum_{e\in \mathbf{\partial}S} w_e,
\end{equation}
where $w_e$ is the edge weight in the case of weighted hypergraphs. 

\subsubsection{Random Walks and Laplacian}
In dyadic graphs, a sequence of vertices is sufficient to define a walk, since there is only one way to traverse from one node to another in one step. However, due to the flexibility of hypergraphs, one has to specify the order of both edges and walks. 

\begin{definition}[$s$-walk]
\label{def8:swalk}
Let $\mathcal{H}$ be an $r$-uniform hypergraph, for $1\leq s \leq r-1$, an $s$-walk of length $k$ is defined as a sequence of vertices $$v_1, v_2, ..., v_j, ..., v_{(r-s)(k-1)+r}$$ together with a sequence of edges 
$F_1,F_2,...,F_k$ such that $$F_i=\{v_{(r-s)(i-1)+1},v_{(r-s)(i-1)+2},...,v_{(r-s)(i-1)+r}\}.$$
\end{definition}

Basically, any two adjacent edges of an $s$-walk have exactly $s$ vertices in their intersection. For the vertex set $V$, let $V^{\underline{s}}$ be the set of all ordered $s$-tuples consisting of $s$ distinct elements in $V$. For example when $s=2$, $$V^{\underline{2}} = \{(v_1,v_2),(v_1,v_3),...,(v_2,v_1),(v_2,v_3),...\}.$$ To compute the Laplacian, we consider the following two cases~\cite{lu2011highordered}:

(1) In the case of $1\leq s \leq r/2$, for any $F_i$, there will not be any intersection with $F_{i+2}$ or $F_{i-2}$. So, the $s$-walk can be interpreted as the walk on a weighted dyadic graph. We define a weighted undirected graph $G^{(s)}$ over $V^{\underline{s}}$ as follows. Let the weight $w(x, y)=|\{F\in \mathcal{E}: [x]\cup [y]\subseteq F\}|$. Here, $[x]\cup[y]$ is the disjoint union of $[x]$ and $[y]$, and $x,y$ are vertices of $s$-tuple of the original vertices. Then, we define the \textit{$s$-th Laplacian} $\mathcal{L}^{(s)}$ of hypergraph $\mathcal{H}$ to be the Laplacian of graph $G^{(s)}$.

(2) Another case is $r/2 < s \leq r-1$, where $F_i$ also intersects with $F_{i+2}$ or $F_{i-2}$ (if it exists). We define a directed graph $D^{(s)}$ over the vertex set $V^{\underline{s}}$ as follows. With $x,y$ are still the $s$-tuples of the original vertices, let $(x, y)$ be a directed edge if $x_{r-s+j}=y_j$ for $1\leq j\leq 2s-r$ and $[x]\cup[y]$ is an edge of $\mathcal{H}$. Then, we define the \textit{$s$-th Laplacian} $\mathcal{L}^{(s)}$ of hypergraph $\mathcal{H}$ to be the Laplacian of Eulerian directed graph $D^{(s)}$. 

\subsubsection{Downgrading a Hypergraph to a Dyadic Graph}
\label{downgrade hypergraph}
To downgrade hypergraphs to dyadic graphs, one of the most straightforward ways is to perform \textit{(clique) expansion}. For each hyperedge $e$, we enumerate all its size-two subedges to form a dyadic graph. Depending on the application, one can either inherit weights from the original hypergraph for these new edges or only keep the structural information~\cite{10031242}. While it is often much more convenient to perform dyadic graph algorithms, the expanded new graph indeed loses higher-order information. 

A more general expansion is the \textit{Multi-Level decomposition}~\cite{inproceedings2020kijung}. In Multi-Level decomposition in addition to enumerating size-two subedges to form a dyadic graph, we construct hypernodes based on the coexistences within the original hyperedges. For example, assume there is a hyperedge of size $|e|$. At each layer $k$, we generate all possible $|e| \choose k$ hypernodes of size $k$ to form a clique. By selecting $k$ ranging from 2 to the maximum order of the hypergraph, we construct a corresponding dyadic graph for each $k$. The most appealing feature of such a decomposition is that one can easily reconstruct the hypergraph from its expansions across all layers.

\subsection{Use Cases and Applications}
Here, we summarize studies utilizing hypergraphs. One branch focuses on developing tools for network analysis, mostly extending graph theories to hypergraph-level. Another branch applies hypergraph to model higher-order interactions for network analysis.  

\subsubsection{Network Measurements}
Many network properties and measurements are generalized from dyadic graphs to hypergraphs. As an analogy to walks on dyadic graphs, $s$-walk is proposed and applied to hypergraphs~\cite{DBLP:journals/epjds/AksoyJMPP20}. As mentioned in Definition \ref{def8:swalk}, an $s$-walk is a series of hyperedges where the intersection nodes between any adjacent hyperedges have size greater than $s$. More specifically, given a larger $s$ indicates a component filled by denser overlapped hyperedges. Consequently, $s$-connected components and $s$-distance are also defined based on $s$-walks, which forms a series of hypergraph analysis tools. Such measurements distinguish real-world networks and random hypergraphs configured from network models. 

Centrality measures assess how important a node is in terms of its position and how it connects to other nodes in the graph. Three eigenvector centralities for uniform hypergraphs are defined by Benson~\cite{DBLP:journals/corr/abs-1807-09644}. Similar to dyadic networks, the centrality of a node in hypergraph can be influenced by centralities of all its neighbors. For each specific hyperedge to which it belongs, there could be a weighting function based on the centrality scores of its neighbors on that edge. First version is called \textit{Clique motif Eigenvector Centrality}, where it simply refers to the total sum of centrality of all neighbors; Second is \textit{$Z$-Eigenvector Centrality}, which multiplies the neighbors' centralities on each hyperedge and then sums them up; Third is \textit{$H$-Eigenvector Centrality}, which is the square root of the \textit{$Z$-Eigenvector Centrality}. Experimental results show that none of these three centralities is consistently superior to others. So, one has to consider a specific objective to make an informed selection.

As motifs are significant subgraph patterns in dyadic graphs, \textit{higher-order motifs} are defined in a similar way~\cite{Lotito_2022}. Given a set of nodes of size $k$, a higher-order motif is formed by a collection of hyperedges consisting of only these $k$ nodes. As $k$ increases, the motif variations can exponentially increase, which makes it impossible to enumerate and detect all motifs in the hypergraph. Addressing this problem, Lee et al.~\cite{DBLP:journals/corr/abs-2003-01853} propose \textit{hypergraph motifs}, which is a special kind of higher-order motif. Hypergraph motifs have a fixed structure that consists of three connected hyperedges. Based on overlapping nodes in hyperedges, all nodes are classified into one of the seven possible areas in a Venn diagram. Such a structure significantly simplifies the process of motif detection and checking for isomorphisms.

\subsubsection{Generative Models}
Graph generation algorithms aim to generate realistic graphs similar to those observed in the real world. Chodrow generalizes two variants (\textit{vertex-labeled} and \textit{stub-labeled}) of the \textit{configuration model}, a well-known graph generation algorithm, to hypergraphs~\cite{10.1093/comnet/cnaa018}. Configuration model in dyadic graphs requires the knowledge of the degree sequence. In hypergraphs, in addition to degree sequences, dimension sequence (sizes of edges) is also needed to generate a random graph. The vertex-labeled hypergraph configuration model is just a uniform distribution over the space of hypergraphs defined by degree and dimension. The stub-labeled hypergraph configuration model simply copies nodes as many times as their degrees and places them into a multiset. The algorithm then uniformly samples hyperedges based on the dimension sequence, where each node can only appear once in a specific hyperedge. 

Lee et al.~\cite{DBLP:journals/corr/abs-2101-07480} extend the \textit{Chung-Lu} model~\cite{doi:10.1073/pnas.252631999} to hypergraphs (\textit{HyperCL}) by ensuring to preserve the distribution from the given degree sequence and the edge-size sequence as input. However, real-world hypergraphs exhibit stronger communities over random graphs. Addressing this issue, the authors further propose \textit{HyperLap}, a multilevel HyperCL that introduces a group parameter $L$ at each level, which aims to help reconstruct community patterns of real-world hypergraphs. New hyperedges generated within each group are expected to have high number of overlapping nodes, especially when the group is small.

Furthermore, a later study extends the \textit{degree-corrected stochastic blockmodel}~\cite{2011Stochastic}, which is a generative model of graphs with both community structure and degree sequences, to hypergraphs~\cite{ doi:10.1126/sciadv.abh1303}. For hyperedge candidates, the authors introduce an affinity function to compute the wiring possibility based on the group memberships of their nodes. Basically, more nodes in the same group have a higher probability of forming hyperedges. Three estimates ---the affinity function, node labels, and node degrees, are alternatively learned by optimizing a likelihood function. To solve this objective, the authors propose an `All-or-Nothing' (AON) \textit{Louvain}-type algorithm~\cite{blondel2008fast} under the assumption that hyperedges are expected to lie fully within the cluster. Experimental results on both synthetic and empirical data validate the efficiency and accuracy of the framework.

\begin{table*}[ht]
\centering
\caption{Tools for Discovering Motifs}
\label{tab7.1:motif_tools}
\adjustbox{max width=\textwidth}{
\begin{tabular}{|l|l|l|l|} \hline
Package Name & Year & Description & Official Link (if exist)\\ \hline
MFinder~\cite{articlemfinder} & 2005& motif detection, enumeration/edge sampling & \url{https://www.weizmann.ac.il/mcb/UriAlon/download/ParTI} \\ \hline
MAVisto~\cite{10.1007/11599128_7} & 2005 & motif detection, enumeration & \url{https://kim25.wwwdns.kim.uni-konstanz.de/vanted/addons/mavisto/} \\ \hline
FANMOD~\cite{10.1109/TCBB.2006.51} & 2005 & motif detection, enumeration/node sampling & \url{https://github.com/gabbage/fanmod-cmd} (unofficial) \\ \hline
Grochow–Kellis~\cite{10.5555/1758222.1758229} & 2007 & motif detection, mapping & \url{https://github.com/jptboy/CSCI3104_GC2} (unofficial) \\ \hline
MODA~\cite{articlemoda} & 2009 & motif detection, mapping/sampling, undirected only & \url{https://github.com/smbadiwe/ParaMODA} (unofficial)\\ \hline
Kavosh~\cite{Kashani2009Kavos-40487} & 2009 & motif detection, enumeration &  \url{https://github.com/shmohammadi86/Kavosh} \\ \hline
G-Tries~\cite{inproceedingsgtrip} & 2010 & motif detection, enumeration/mapping, undirected only & \url{https://www.dcc.fc.up.pt/gtries/}\\ \hline
TemporalMotif~\cite{DBLP:journals/corr/ParanjapeBL16} & 2016 & temporal motif count &
\url{http://snap.stanford.edu/temporal-motifs/}\\ \hline
MODET~\cite{articlemodet} & 2019 & motif detection, mapping, undirected only & \url{https://github.com/sabyasachipatra/modet}\\ \hline
\end{tabular}
}
\end{table*}

\begin{table*}[ht]
\centering
\caption{{Tools for Learning Simplicial Complexes}}
\label{tab7.2:sc_tools}
\adjustbox{max width=\textwidth}{
\begin{tabular}{|l|l|l|l|} \hline
Package Name & Environment & Description & Official Link\\ \hline
Simplicial & Python & topology, homology, filtrations & \url{https://simplicial.readthedocs.io/en/latest/}\\ \hline 
Javaplex~\cite{Javaplex} & Matlab/Java & persistent homology, filtrations & \url{https://github.com/appliedtopology/javaplex}\\ \hline 
Ripser~\cite{Bauer2021Ripser} & C++ & persistent homology, Vietoris–Rips filtrations & \url{https://github.com/Ripser/ripser}\\ \hline 
simplextree & R & topology &\url{https://github.com/peekxc/simplextree}\\ \hline 
Simplicial.jl & Julia & simplicial complexes, directed complexes &\url{https://github.com/nebneuron/Simplicial.jl}\\ \hline 
simplicial-complex & JavaScript & structural and topological operations& \url{https://www.npmjs.com/package/simplicial-complex}\\ \hline 
Dionysus 2 & C++ & persistent homology & \url{https://mrzv.org/software/dionysus2/}\\ \hline 
DIPHA & C++ & distributed, persistent homology & \url{https://github.com/DIPHA/dipha}\\ \hline 
Perseus~\cite{10.1007/s00454-013-9529-6} & C++ & persistent homology & \url{https://people.maths.ox.ac.uk/nanda/perseus/}\\ \hline 
Moise & Maple & homology groups &\url{https://www.math.drexel.edu/~ahicks/Moise/}\\ \hline
TopoEmbedX~\cite{hajij2023topological} & Python & representation learning &\url{https://github.com/pyt-team/TopoEmbedX}\\ \hline
\end{tabular}
}
\end{table*}

\subsubsection{Hypergraph Partitioning and Clustering}
Hypergraph partitioning methods are generalized from classical graph cut problems. Veldt et al.~\cite{DBLP:journals/corr/abs-2001-02817} propose a comprehensive set of steps for solving hypergraph $s-t$ cuts problem. The first step is to select a splitting function, which maps the hyperedge that is going to be cut to a real number penalty (this has to be defined specifically). The authors specify a property for the splitting functions called \textit{cardinality-based}, where the penalty only correlates with the sizes of split clusters. The hypergraph $s-t$ cuts problem is defined as minimizing the total splitting penalty for all crossing hyperedges. Various splitting functions are analyzed and tested over real-world datasets. Based on the results, a new clustering framework for hypergraphs is proposed~\cite{10.1145/3394486.3403222}, which minimizes the localized \textit{ratio cut} objective. The algorithm requires a set of input nodes and returns a well-connected cluster that highly overlaps with the inputs. The running time of this algorithm only depends on the size of input set, and guarantees cuts or conductance under a specific bound. 

The hypergraph cut problem can also be solved with tensor representations. By extending matrix-based methods, a tensor spectral clustering method is developed for partitioning higher-order networks~\cite{DBLP:journals/corr/BensonGL15}. For example, an order-3 undirected network can be represented as an order-3 symmetric tensor. As an analog to random walk on dyadic networks, a second-order Markov process is applied to express state changes on order-3 networks. Clustering of higher-order networks can be achieved by recursively partitioning the graph by minimizing \textit{sweep cuts}. Such a clustering method preserves higher-order structures rather than just edges, as shown through experiments on both synthetic and real-world networks. 

The clustering method can be further extended when introducing additional information. As for labeled networks, Amburg et al.~\cite{10.1145/3366423.3380152} propose a novel hypergraph clustering framework based on given edge labels. The objective simultaneously minimizes (1) edges across clusters and (2) edges that do not belong to the assigned cluster. These two requirements can be combined by simply counting the number of nodes whose labels are inconsistent with their connected edges. In the case of two categories, this problem reduces to an s-t cut problem by forming a dyadic graph and adding a terminal node to it. In the case of more than two categories, multiple approximation algorithms for such  an NP-hard problem are developed, such as an LP relaxation and multiway cuts. Experimental results on synthetic and real-world graphs show that the proposed method outperforms baselines including \textit{Majority Vote}, \textit{Chromatic Balls} and \textit{Lazy Chromatic Balls}. 

As discussed in Section~\ref{downgrade hypergraph}, downgrading hypergraph to dyadic is also an effective way to perform existing algorithms. Liu et al.~\cite{osti_10285904} propose a hypergraph clustering method, called \textit{Local Hypergraph Quadratic Diffusions (LHQD)}. The first step of LHQD reduces the hypergraph to a directed graph that preserves the conductance property of the original graph. The equality of conductance is achieved by introducing auxiliary nodes for each node. The second step of LHQD creates a source and a sink node in the directed graph, whose weights to the auxiliary nodes are equal to their degrees. Such a conceptual transformation ensures that the objective function becomes the same as the original problem. The performance is validated by performing clustering on two real-world networks. 

\subsubsection{Modeling Higher-Order Interactions}
\label{sec:hypergraph-modeling}
Many real-world interactions can be modeled directly as hypergraphs. The entities often have different types, but in terms of hypergraphs, research rarely emphasizes on node heterogeneity.  
Tan et al.~\cite{10.1145/2037676.2037679} model the music recommendation problem as a hypergraph ranking problem. At the beginning, different objects (users, groups, tags, tracks, albums, and artists) are represented as nodes and pairwise relationships among them are identified. Hyperedges are created by combining edges based on the intrinsic connections among them, e.g. tracks in the same album. Given a query (set of nodes), the recommendation is based on a ranking of scores of all other nodes on the hypergraph. This scoring process is trained by the ground truth of node labels under the smoothing constraint, which close nodes should have similar scores. 

Similar to music recommender systems, the image retrieval problem can be modeled as the hypergraph ranking problem, where images are nodes that are assigned to hyperedges based on similarities. Liu et al.~\cite{LIU20112255} applied a \textit{soft hypergraph} model in which the entries on the incidence matrix are calculated using some similarity functions instead of arbitrary values. Hyperedges are created by selecting any node as centroid and adding its $k$ nearest neighbors. When an image query comes, it solves the linear system based on a cost function of hypergraph partition problem, which ensures vertices sharing many incidental hyperedges to obtain similar labels. 

Rather than built on the basis of similarities, hypergraphs are also directly used for modeling data in biology. Patro and Kingsford~\cite{10.1093/bioinformatics/btt224} model \textit{network history inference} using a hypergraph structure. Generally speaking, \textit{network history inference} is to find a small set of tuples that record the historical interactions between leafs on the protein network. The mapping of different states is represented as a hypergraph, where current state and correlated historical states are connected by order-3 hyperedges. The problem is solved by minimizing total cost over the network. Such model can be applied to reconstruct the ancestral networks or to predict missing links. 

\subsubsection{Hypergraph Neural Networks}
Graph neural networks have been generalized beyond pairwise interactions in modeled as hypergraphs. Hypergraphs can be applied to either (1) model higher-order graph data as input matrices, such as the incident matrix of hypergraphs; or to (2)build multilayer neural network structures but using higher-order forward- and back-propagation instead. Many state-of-the-art models, especially for graph neural networks, are extended to hypergraphs. Examples include hyper- models such as
HGNN~\cite{https://doi.org/10.48550/arxiv.1809.09401},
HGAT~\cite{DBLP:journals/corr/abs-1901-08150,DBLP:journals/corr/abs-2112-14266},
MHCN~\cite{DBLP:journals/corr/abs-2101-06448}, and
HGCN~\cite{https://doi.org/10.48550/arxiv.1809.02589,10.1145/3397271.3401133}. 
For a comprehensive survey on graph neural networks, readers are referred to the survey by Thomas et al.~\cite{gnnsurvey2022}. 

\begin{table*}[ht]
\centering
\caption{Tools for Learning Hypergraphs}
\label{tab7.3:hypergraph_tools}
\adjustbox{max width=\textwidth}{
\begin{tabular}{|l|l|l|l|} \hline
Package Name & Environment & Description & Official Link \\ \hline
HyperG & R & Hypergraph Modeling & \url{https://cran.r-project.org/web/packages/HyperG/} \\ \hline
HyperNetX~\cite{DBLP:journals/corr/abs-2003-11782} & Python & Hypergraph Modeling, Visualization &\url{https://github.com/pnnl/HyperNetX} \\ \hline
GraphML~\cite{Brandes2013GraphML} & XML & File Format & \url{http://graphml.graphdrawing.org/index.html} \\\hline
hypergraph & R & Hypergraph Modeling & \url{https://bioconductor.org/packages/3.15/bioc/html/hypergraph.html} \\ \hline
SimpleHypergraphs.jl~\cite{Spagnuolo_2020}& Julia & Hypergraph Modeling, Visualization & \url{https://github.com/pszufe/SimpleHypergraphs.jl} \\ \hline
halp & Python & Hypergraph Modeling, Algorithms&\url{https://murali-group.github.io/halp/} \\ \hline
kahypar~\cite{DBLP:phd/dnb/Schlag20} & Python &Hypergraph Partitioning& \url{https://pypi.org/project/kahypar/}\\ \hline
Tensorly~\cite{JMLR:v20:18-277} & Python & Tensor Learning &\url{http://tensorly.org/stable/index.html} \\
\hline
Tensors.jl~\cite{articletensorjl} & Julia & Tensor Learning & \url{https://juliahub.com/ui/Packages/Tensors/F7rKl/1.11.0} \\ \hline
rTensor~\cite{JSSv087i10} & R & Tensor Learning & \url{https://cran.r-project.org/web/packages/rTensor} \\ \hline
\end{tabular}
}
\end{table*}

\begin{table*}[ht]
\centering
\caption{{Applications of Higher-Order Networks}}
\label{tab8:summasion}
\adjustbox{max width=\textwidth}{
\begin{tabular}{|l|c|c|c|} \hline
 & Network Motifs & Simplicial Complexes & Hypergraphs \\ \hline
Statistical Significance & \cite{ShenOrr2002NetworkMI}\cite{Kashtan13773}\cite{brain_article}\cite{Duclos2020BrainNM}\cite{articleclique} &  & \\ \hline
Graph Classification & \cite{alon-network-motifs}\cite{doi:10.1126/science.1089167} &  & \\ \hline
Network Modeling & \cite{10.1093/bioinformatics/bth436}\cite{leskovec2010signed} & \cite{Courtney_2016}\cite{Young_2017}\cite{2019simplicialm}\cite{Bianconi_2017} & \cite{10.1093/comnet/cnaa018}\cite{DBLP:journals/corr/abs-2101-07480}\cite{doi:10.1126/sciadv.abh1303}\\ \hline
Clustering & \cite{doi:10.1126/science.aad9029}\cite{10.1145/3097983.3098069}\cite{2018honhao}\cite{doi:10.1126/science.aad9029}\cite{DBLP:journals/corr/abs-1802-06916} & \cite{DBLP:journals/corr/abs-1708-08436} & \cite{DBLP:journals/corr/abs-2001-02817}\cite{10.1145/3394486.3403222}\cite{DBLP:journals/corr/BensonGL15}\cite{10.1145/3366423.3380152}\cite{osti_10285904}\\ \hline
Representation Learning & \cite{rossi2018hone}\cite{10.1145/3473911} & \cite{DBLP:journals/corr/abs-2103-04046}& \cite{8999197}\cite{gong2023generative} \\ \hline
Link Prediction & \cite{DBLP:journals/corr/abs-1902-06679} & \cite{DBLP:journals/corr/abs-1802-06916} & \cite{10.1093/bioinformatics/btt224}\cite{8999197}\\ \hline
Persistent Homology &  & \cite{Ghrist2005Coverage}\cite{De_coveragein}\cite{5444998}\cite{Nanda2014SimplicialMA}\cite{10.1093/bioinformatics/btm033}\cite{DeWoskin2010ApplicationsOC}\cite{DBLP:journals/aaecc/ArsuagaBDMPP12}\cite{doi:10.1098/rsif.2014.0873}\cite{doi:10.1073/pnas.1506407112}\cite{https://doi.org/10.48550/arxiv.1601.01704}\cite{10.1145/3274895.3274952} & \\ \hline
Analysis Tools and Measurements & \cite{2018honhao}\cite{doi:10.1126/science.aad9029} & \cite{DBLP:journals/corr/abs-1807-05044} & \cite{DBLP:journals/epjds/AksoyJMPP20}\cite{DBLP:journals/corr/abs-1807-09644}\cite{Lotito_2022}\cite{DBLP:journals/corr/abs-2003-01853}\\ \hline
Recommender System &  &  & \cite{10.1145/2037676.2037679}\cite{LIU20112255}\\ \hline
Neural Networks &  &  &\cite{https://doi.org/10.48550/arxiv.1809.09401}\cite{DBLP:journals/corr/abs-1901-08150}\cite{DBLP:journals/corr/abs-2112-14266}\cite{DBLP:journals/corr/abs-2101-06448}\cite{https://doi.org/10.48550/arxiv.1809.02589}\cite{10.1145/3397271.3401133} \\ \hline
\end{tabular}
}
\end{table*}

\section{Datasets and Tools}
\label{sec:dataset_tool}
We summarize some available datasets and tools for studying higher-order networks. As some of these official links might disappear in the future, we provide a comprehensive backup of all tools and datasets in our own repository.\footnote{\url{https://github.com/haotian-syr/HON-tools}}

\subsection{Higher-Order Network Tools}
For most network motif based studies, the first step is to find motifs. In Table~\ref{tab7.1:motif_tools}, we list some scalable algorithms for enumerating or counting motifs. Among these methods, \textit{enumeration} indicates an exhaustive search through the whole graph. \textit{Sampling} indicates that the method calculates an estimated frequency of a given motif by sampling the node/edge and exploring its neighborhood. The \textit{mapping} strategy is a reverse process of enumeration, which maps the given motif onto the whole network. 

Table~\ref{tab7.2:sc_tools} collects packages and software for studying simplicial complex. Some are tagged as `topology', which are comprehensive packages that build the data structure from the lower level information. Some are software that are easy-to-use for most popular applications such as persistent homology and filtrations.

Table~\ref{tab7.3:hypergraph_tools} collects packages for modeling hypergraphs and tensors. Most packages provide a data structure and implement the most basic algorithms using it; some packages support network visualization.

\subsection{{Higher-Order Datasets}}
We summarize some dataset resources with higher-order interactions:
\begin{itemize}
    \item ARB Data\footnote{\url{https://www.cs.cornell.edu/~arb/data/}}: A dataset repository (19 datasets--4 with millions of nodes, 10 with thousands of nodes, and 5 with hundreds of nodes) collected by Austin R. Benson. Most are higher-order networks from various fields, including temporal and labeled hypergraphs. 
    \item LINQS\footnote{\url{https://linqs.soe.ucsc.edu/}}: A collection of 11 relational datasets (1 with a million nodes, others are thousands of nodes or less). Many of them are collaboration networks, which naturally include higher-order interactions.
    \item Twitter Data\footnote{\url{https://data.world/datasets/twitter}}: Tweets can be modeled as higher-order networks by taking hashtags as nodes and co-appearances as edges (require preprocessing). Besides these, one can search twitter datasets online for any specific interest or collect data using APIs. 
    \item Temporal Co-authorship~\footnote{\url{https://github.com/kswoo97/pcl}}: Three large-scale (sizes: 27 million / 13 million / 41 thousand nodes) hypergraph datasets in both static and temporal forms~\cite{articledataset}. 
    
\end{itemize}

\subsection{{Expected Time Complexities}}
While time complexity of exploring higher-order networks can vary across topologies and algorithms, some time complexities are typically expected for some basic higher-order algorithms. Here, we briefly list some expected time complexities for analyzing higher-order networks. \vspace{1mm}

\noindent \textbf{Motif counting}: The time complexity of counting motifs depends highly on the structural complexity of given motifs as the essential algorithm for finding motifs involves checking for subgraph isomorphism, which is known to be NP-complete. Most fast motif-finding algorithms focus on size 3 or 4 motifs. For example, counting motifs of size 3 (triangles) can be solved in $O(|E|d_{max})$~\cite{7373304} and counting motifs of size 4 can be solved in $O(|E|d_{max}+|E|^2)$~\cite{MARCUS2012810}, where $d_{max}$ is the maximum degree in the graph. 

\noindent \textbf{Homology groups}: The time complexity of computing homology groups of the simiplical complex is $O(n^{\omega})$, where $n$ is the number of simplices and the exponent $\omega \leq 2.4$~\cite{10.1145/1998196.1998229}. Such an acceptable and stable complexity facilitates the wide usage of homology methods. 


\section{Conclusions and Future Directions}
\label{sec:conclusion}
We survey essential algorithms and applications in the literature of higher-order network modeling and analysis. In Table~\ref{tab8:summasion}, we summarize the applications collected in this survey. Due to the scope of this scope, we have highlighted representative studies from each field. To explore each area comprehensively, we have directed readers to other surveys focusing on each domain.

However, research on higher-order networks has significant future potential. Here we list some open problems or under-explored research directions:

\subsection{Data Source and Modeling}
\noindent Unlike dyadic graphs, not many repositories of higher-order network data are available. Building tools to collect, store, and model higher-order data is of significant interest for various academic use cases. Below we list some essential, yet under developed, tools for studying higher-order data. 

\vspace{1mm}
\noindent \textbf{\underline{Recovering Higher-order Data in Dyadic Graphs}}: \\Most existing network data is collected in dyadic form, which has already lost higher-order interactions. In motif-based studies, one cannot distinguish whether a specific motif is indeed a higher-order interaction or formed by a combination of dyadic interactions. It is therefore a challenge worth addressing to build tools that can distinguish higher-order interaction or that can rebuild higher-order networks from a dyadic graph. 

\vspace{1mm}
\noindent \textbf{\underline{Dynamic Higher-Order Graph}}: Most higher-order interactions are associated with time, such as protein interactions, hashtags, and group chats. Rather than a snapshot analysis of a network during some small interval, there is a demand for tools that can maintain and access the whole or partial network structure of a temporal higher-order network. Such tools enable real-time fast algorithms for various tasks including link prediction, community detection, anomaly detection, and the like.  

\vspace{1mm}
\noindent \textbf{\underline{Matrix/Tensor Representation}}: Topologies representing higher-order interactions are always associated with matrix or tensor representations, such as motif matrix~\cite{doi:10.1126/science.aad9029}, tensors~\cite{DBLP:journals/corr/BensonGL15} and incidence matrix. However, due to complexity and lack of mathematical support, algorithms on these matrix representations are not explored as extensively as matrix-based methods for dyadic graphs. 

\vspace{1mm}
\noindent \textbf{\underline{{Interpretability and Causality}}}: Most higher-order networks studies develop algorithms and applications that explore collected higher-order data. Current research rarely aims to interpret findings in higher-order networks or understand why some high-order interactions or patterns exist. As mentioned, some patterns might reflect important real-world information beyond network structures, such as the small functional unit in brain networks~\cite{brain_article}. Interpretability is especially crucial as more black-box techniques (e.g.,  deep neural networks) or embedding methods are designed for higher-order networks.

\subsection{Machine Learning Applications}
\noindent Many real-world applications have focused on higher-order graphs. Below, we list three general application domains for higher-order networks that have a significant potential for future research.

\vspace{1mm}
\noindent \textbf{\underline{Representation Learning}}: Representation learning is a powerful tool for transforming high-dimensional data into fixed-size vectors and has been extremely successful for downstream machine learning tasks, such as node classification, link prediction, among others. However, not many representation learning methods are introduced for higher-order networks. Hence, there is a significant  demand for representation learning methods that can embed higher-order graphs both at the node-level~\cite{8999197,gong2023generative} and the graph-level.

\vspace{1mm}
\noindent \textbf{\underline{Recommender Systems}}: One of the most direct uses of higher-order networks is in recommender systems. For example, nodes involved in same hyperedge can share some similarities. As discussed in Section \ref{sec:hypergraph-modeling}, some applications such as music recommendation~\cite{10.1145/2037676.2037679} and image retrieval~\cite{LIU20112255} are developed. However, for individual recommendations, the advantage of higher-order graphs over simple or heterogeneous graphs needs to be further studied. Similarly, group recommendation~\cite{10.14778/1687627.1687713} is another direction that has the potential to be studied. 

\vspace{1mm}
\noindent \textbf{\underline{Graph Neural Networks}}: Similar to modeling real-world interactions, the structure of neural networks can be designed to accept higher-order interactions as input when necessary. Due to insufficient studies on hypergraphs, there is only limited work that directly utilizes hypergraphs in neural networks. One main challenge is to extend the adjacency matrix, where some studies have considered the incidence matrix as a solution to this challenge~\cite{https://doi.org/10.48550/arxiv.1809.09401}. 

%

\bibliographystyle{unsrtnat}
\bibliography{reference}  






\end{document}